\documentstyle[12pt,epsf]{article}
\newcommand{\rf}[1]{(\ref{#1})}
\newcommand{\beq}{\begin{equation}}
\newcommand{\eeq}{\end{equation}}
\newcommand{\bdm}{\begin{displaymath}}
\newcommand{\edm}{\end{displaymath}}
\newcommand{\bea}{\begin{eqnarray}}
\newcommand{\eea}{\end{eqnarray}}
\newcommand{\nn}{\nonumber \\}

\newcommand{\pa}{\partial}

\def\sepand{\rule{14cm}{0pt}\and}
\newcommand{\la}{\langle}
\newcommand{\ra}{\rangle}

\def\gsim{\raise.3ex\hbox{$>$\kern-.75em\lower1ex\hbox{$\sim$}}}
\def\lsim{\raise.3ex\hbox{$<$\kern-.75em\lower1ex\hbox{$\sim$}}}
\def\sepand{\rule{10cm}{0pt}\and}
\def\vek#1{{\bf #1}}
\begin{document}
\topmargin 0cm
\topskip 0mm
 {\title
{\null\vskip-3truecm
{ \hskip10truecm {\small NORDITA-94/23 P\hfill }\vskip 0.1cm}
%{ \hskip10truecm {\small hep-ph/xxxxxx\hfill }\vskip 1.5cm}
{\bf On the evolution of magnetic fields in the early universe}}
\author{
{\sc Kari Enqvist$^1$} \\
{\sl Nordita} \\
{\sl Blegdamsvej 17,
DK-2100 Copenhagen, Denmark}\\
and\\
{\sc Poul Olesen$^2$ }\\
{\sl The Niels Bohr Institute, University of Copenhagen} \\
{\sl Blegdamsvej 17, DK-2100 Copenhagen, Denmark} \\
\sepand
}\maketitle}

% \vspace{1.5cm}
\begin{abstract}
\noindent
We address the question of the existence of magnetic domains in a recent
scenario for the creation of primordial magnetic fields. We write down
the magnetohydrodynamic equation in an expanding universe in the infinite
conductivity limit and discuss its solution. No domains are found.
\end{abstract}
\vfil
\footnoterule
{\small  $^1$enqvist@nbivax.nbi.dk;  $^2$polesen@nbivax.nbi.dk}
\thispagestyle{empty}
\newpage
\setcounter{page}{1}
%%%%%%%%%%%%%%%%%%%%%%%%%%%%%%%%%%%%%%%%%%%%%%%%%%%%%%%%%%%%%
We have recently presented a scenario for the creation of a primordial magnetic
field at the GUT scale \cite{us}. This field is due to nonperturbative
quantum effects in the early universe. For a typical realistic GUT such as the
supersymmetric SU(5) model we found a field of the order $B\simeq 5\times
10^{-8}\mu^2$ where $\mu$ is the renormalization point with $\mu\simeq
T_{\rm GUT}\simeq 10^{15}$ GeV.

Once the field is imprinted on the charged plasma it will remain because
the early universe is a very efficient conductor with conductivity inversely
proportional to the collision cross section. For details we refer to the
Appendix of Turner and Widrow \cite{turner}, where the conductivity has
been estimated. It then follows that whatever magnetic flux there existed
in the early universe, it is frozen in the plasma. Diffusion begins
relatively late and starts from small scales.

In order to estimate the magnetic field today one needs to take into account
the fact that in the universe there exists causal horizons. Naively one could
argue that each horizon "bubble" (i.e. a bubble with a radius equal to the
causal horizon at the GUT time $ H^{-1}\simeq 10^{-28}$ cm) constitutes a
magnetic domain. Thus one should average over all random domains, which
now have a size of about 1 m, in order to obtain the average magnetic field
today. If this were so, then the value of $B$ at the time of protogalaxy
collapse would very likely to be too small to be of interest for the
observed galactic magnetic fields.

In the present note we shall, however, show that the picture of a causal
horizon bubble as a magnetic domain with randomly pointed field, such as
in a ferromagnet, is not correct. Instead the proper physical situation
is that when two bubbles meet, a new "joint" bubble is formed with the
magnetic field lines frozen to the plasma particles. After a short time
the joint plasma looses all memory of its origin, and so does the magnetic
field, too. The situation is analogous to having two containers filled
with identical gases which are separated by a  wall. Once the
wall is removed the joint system has no memory of the original two separate
containers.

When the electric conductivity is very large the only electric field that
can exist is induced by the motion of   charged particles in the magnetic
field.
Here it is very likely essential that the field $B$ is very small relative
to the scale $\mu^2\simeq T^2$,
as is the case in our ferromagnetic universe model
\cite{us}. If the field is large, say of the order of
$T^2$, the mean free path of the charged plasma particles is large in
comparision with their radius of curvature in the magnetic field. In such a
case  conductivity cannot be estimated from the drift of the charged particles
and in general it depends on the magnetic field itself.

We use the same conventions as in ref. \cite{turner}. In particular, we adopt
the conformal metric
\beq
d\tau^2=a(\eta)^2(-d\eta^2+dx^2+dy^2+dz^2)~,
\label{metric}
\eeq
where $\eta$ is the conformal time. The advantage of the conformal metric is
that electric and magnetic fields a treated on an equal footing with
\beq
F_{12}=a^2B_3~~({\rm and~~ cyclic}),~~~F_{01}=-a^2E_1,~~ ({\rm etc.})
\label{fmunu}
\eeq

The standard cosmological model described by Eq. \rf{metric} is now assumed to
be valid {\sl on the average.} We treat the metric as a background field
not affected by the presence of the magnetic field itself. In the relation
physical length scale = $a\times$(comoving length scale) the comoving
length scale related to a charged plasma particle is assumed to be fixed
(as in the standard cosmological scenario) only on the average. Similarly,
the charged particles are only assumed to be at rest at a given
Friedman-Robertson-Walker time in an average sense. Thus, the particles are
associated with a velocity field ${\bf v}(\vek x,t)$.
With the definitions \rf{fmunu} one of the Maxwell equations reads
\beq
{\pa a^2\vek B\over\pa\eta}=-\nabla\times a^2\vek E~,
\label{maxwell}
\eeq
where the spatial derivative is with respect to the comoving coordinates.
Now, because of the very large conductivity $\vek E$ must be an induced
field, so that
\beq
\vek E=-\vek v\times\vek B~.
\label{lorentz}
\eeq
Hence Eq. \rf{maxwell} becomes
\beq
{\pa a^2\vek B\over\pa\eta}=\nabla\times(\vek v\times a^2\vek B)~,
\label{mhd}
\eeq
which is the magnetohydrodynamics equation for infinite conductivity
(see \cite{dynamo} for discussion of this subject) in an expanding
universe.

Eq. \rf{mhd} can be written as
\beq
{\pa a^2\vek B\over\pa\eta}=a^2(\vek B\cdot\nabla)\vek v-(\vek v\cdot\nabla)
a^2\vek B-a^2\vek B\nabla\cdot\vek v~.
\label{rewrite}
\eeq
In order to proceed we need to know the divergence of $\vek v$. In the
nonrelativistic limit (ignoring $a$) it is given by the usual continuity
equation
\beq
{\pa n\over\pa t}+\nabla\cdot (n\vek v)=0.
\label{cont}
\eeq
In general relativity Eq. \rf{cont} is replaced by
\beq
(nU^\mu);_\mu=U^\mu{\pa n\over \pa x^\mu}+{n\over\sqrt{g}}{\pa\over\pa x^\mu}
(\sqrt{g}U^\mu)=0,
\label{grcont}
\eeq
where $n$ is the density of charged particles which is conserved, and $U^\mu$
is the four-velocity. In the conformal metric we obtain
\beq
-\nabla\cdot\vek v= {1\over a^4U^0n}{\pa\over\pa \eta}(a^4U^0n)
+ {1\over U^0n}\vek v\cdot \nabla nU^0~.
\label{kaava}
\eeq
Inserting this in Eq. \rf{rewrite} we obtain
\bea
{d\over d\eta}\left({a^2\vek B\over a^4U^0n}\right)&\equiv&
\left({\pa\over\pa\eta}+\vek v\cdot\nabla\right){a^2\vek B\over a^4U^0n}\nn
&=&\left({a^2\vek B\cdot\nabla\over a^4U^0n}\right)\vek v~.
\label{totder}
\eea
in an adiabatically expanding universe $n\sim a^{-3}$. To find how $U^0$
scales, let us note that in the FRW metric $U^0_{\rm RW}=dt/d\tau=1$,
so that in the conformal metric $U^0=d\eta/d\tau=1/a(\eta)$. Of course,
in our case this result is valid in the average only, and $\vek v=
\vek U/U^0$ is locally non-vanishing (although $\la\vek v\ra=0$).
Thus we find that
\beq
a^4nU^0\simeq const.
\label{const}
\eeq

Eq. \rf{totder} can be solved. Let $\vek X(\eta_1)$ be the position of a
plasma element at the time $\eta_1$, and let $\vek x=\vek x(\vek X,\eta_2)$
be the position of the same plasma element at the time $\eta_2$. Then
Eq. \rf{totder} is solved by
\beq
{a(\eta_2)^2\vek B_i(\vek x,\eta_2)\over a(\eta_2)^4n(\eta_2)U^0(\eta_2)}
={a(\eta_1)^2\vek B_j(\vek x,\eta_1)
\over a(\eta_1)^4n(\eta_1)U^0(\eta_1)}{\pa x_i\over \pa X_j}~.
\label{sol}
\eeq
Because of Eq. \rf{const} we see that $a^2\vek B$ changes by the factor
${\pa x_i/\pa X_j}$.

Another way of expressing the solution \rf{sol} is the following. Let $\delta
x_i$ be an infinitesimal displacement vector carried by the fluid (in
comoving coordinates). If initially $\delta x_i(0)$ is parallel to
$a^2(0)B_i(0)$,
\beq
 \delta x_i(0)=ca(0)^2B_i(0)~,
\label{deltax}
\eeq
with $c$ a constant, then the solution \rf{sol} implies that the two vectors
$\delta \vek x$ and $a^2\vek B$ move together with the plasma, and at later
times
\beq
 \delta \vek x(\eta)=ca(\eta)^2\vek B(\eta)~
\label{deltaxx}
\eeq
In the standard cosmology the comoving displacement vectors satisfy
\beq
\la\delta x_i(0)\ra=\la\delta x_i(\eta)\ra~,
\label{disp}
\eeq
and hence $a(\eta)^2B_i(\eta)$ is a constant. This means that the magnetic
flux measured in the physical coordinates is constant, which is
the well known result.

Let us  consider $B^2_i$, using the special solution
\rf{deltax}--\rf{deltaxx}. It is clear that
\beq
a(\eta)^4\vek B^2(\eta)=a(\eta)^4\vek B^2(\vek x(\eta),\eta)
=a(0)^4\vek B^2(\vek x(0),0)
\label{clear}
\eeq
statistically. Eq. \rf{clear} is identical to what happens for e.g. the
particle density, for which one finds
\beq
a(\eta)^3n(\vek x(\eta),\eta)=a(0)^3n(\vek x(0),0)~.
\label{density}
\eeq

Now, returning back to the original problem, namely whether different GUT
horizon bubbles form domains or not, we observe that $a^4\vek B^2$ behaves
like the particle density $a^3n$. Of course, the density $a^3n$ does not
have any domains. If two horizon bubbles collide, a new joint bubble is
simply formed, and the information about the original bubbles is lost
(after all, in equilibrium entropy is maximized). Exactly the same thing
happens with respect to $a^4\vek B^2$. If in the initial set-up there are
magnetic domains, these are rapidly washed out by the motion of the
plasma (gas).

So far we have used the special solution \rf{deltax}--\rf{deltaxx}. However,
we may also take the general solution \rf{sol}, which by virtue of
Eq. \rf{const} can be written as
\beq
 a(\eta_2)^2\vek B_i(\vek x(\eta_2),\eta_2)=
 a(\eta_1)^2\vek B_j(\vek x(\eta_1),\eta_1){\pa x_i\over \pa X_j}~.
\label{ratkaisu}
\eeq
{}From a statistical point of view it is natural to assume isotropy so that
\beq
\la{\pa x_i\over \pa X_j}\ra =\delta_{ij}~.
\label{iso}
\eeq
In other words, if we have an original displacement $\delta X_j$ at
$\eta=\eta_1$, then at a later  time $\eta_2$ the corresponding
displacement $\delta x_i$ can in principle go in any direction. However,
the original $j$-direction is the only preferred direction, so that if
$j$ turns into any $i\ne j$ then it is equally likely that the resulting
displacement is positive or negative. Hence the statistical sum is zero.
Eq. \rf{iso} is of course also valid when we consider the standard cosmology
as a statistical model.

Thus Eqs. \rf{ratkaisu} and \rf{iso} again lead us to the conclusion that
$a^4\vek B^2$ is (statistically) constant. They also give rise to the
strong result that
\beq
a(\eta_2)^2\la \vek B_i(\vek x(\eta_2),\eta_2)\ra=
 a(\eta_1)^2\la\vek B_j(\vek x(\eta_1),\eta_1)\ra~,
\label{strong}
\eeq
which demonstrates that, apart from a well  understood redshift factor,
the field $\vek B$ is statistically constant. Eq. \rf{strong}
shows that the evolution of $a^2\vek B$ is {\sl exactly} the same as
the evolution of an invariant density such as $a^3n$ (see Eq. \rf{density})
having no domains. The same result holds for $a^4\la \vek B^2\ra$. It
follows that the magnetic energy $E_B$ scales like $1/a(\eta)$ and hence
$E_Ba$ remains a constant during the course of the evolution of the
universe. Consequently the magnetic energy always corresponds to a
magnetic field of $10^{-8}T^2$. Physically this is so because  in the
infinite conductivity limit there is no dissipation.

We have thus made it plausible that in the framework of the standard cosmology
considered in a statistical sense there are no magnetic domains. Therefore
one may estimate the magnitude of the primordial magnetic field today as
was done in our previous paper \cite{us} with the result that
\beq
B^2_{\rm today}\simeq (10^{-14}{\rm G})^2~.
\label{result}
\eeq
This certainly has the right order of magnitude to act as the seed field
for the galactic dynamo mechanism.

Let us end by reiterating that our argument very much depends on the
fact that in our case the primordial magnetic field is relatively weak.
If the $B$-field is strong, of the order of \cal O$(T^2)$, conductivity
would depend on $\vek B$ and simple magnetohydrodynamics would no longer be
valid. In the construction of Vachaspati \cite{vachaspati}, relying on
the fluctuations of the Higgs fields at the electroweak phase transition,
the field is locally $10^8$ times larger than in our present model.
Hence it is quite possible that such a strong magnetic field does not
freeze in the plasma and domains may exist in that case.

\vskip1truecm\noindent
{\Large\bf Acknowledgements}
\vskip0.5truecm
We thank Tanmay Vachaspati and Victor Semikoz for several hot discussions.

\end{document}